\documentclass[preprint]{JHEP3}
\usepackage{amsmath,amssymb}
\usepackage{stmaryrd}

\newcommand\fverb{\setbox\fverbbox=\hbox\bgroup\verb}
\newcommand\fverbdo{\egroup\medskip\noindent%
\fbox{\unhbox\fverbbox}\ }
\newcommand\fverbit{\egroup\item[\fbox{\unhbox\fverbbox}]}
\newbox\fverbbox

\newcommand{\nlin}[2]{\href{http://xxx.lanl.gov/abs/nlin/#2}{\tt nlin.#1/#2}}

\title{Yangian symmetry of the Y=0 maximal giant graviton}

\author{Niall MacKay$^a\,$\thanks{Email: nm15@york.ac.uk}
~and Vidas Regelskis$^{a\,b\,}$\thanks{Email: vr509@york.ac.uk}\\
$^a$ Department of Mathematics, University of York,\\
$\;\;$ Heslington, York YO10 5DD, UK\\\\
$^b$ Vilnius University Institute of Theoretical Physics and Astronomy,\\
$\;\;$ Go\v{s}tauto 12, Vilnius 01108, Lithuania
}

\preprint{}
\abstract{We study the remnants of Yangian symmetry
of AdS/CFT magnons reflecting from boundaries with no degrees of freedom.
We present the generalized twisted
boundary Yangian of open strings ending on  boundaries
which preserve only a subalgebra $\mathfrak{h}$ of the bulk
algebra $\mathfrak{g}$, where
$(\mathfrak{g},\mathfrak{h})$ is a symmetric pair. This is realized
by open strings ending on the $D3$ brane
known as the $Y=0$ maximal giant graviton in $AdS_5 \times S^5$.
We also consider the Yangian symmetry of the
boundary which preserves an $\mathfrak{su}(1|2)$ subalgebra only.}

\begin{document}

\section{Introduction}

Since the discovery of integrable structures \cite{Zarembo1,Bena,Dolan}
in the AdS/CFT correspondence \cite{Maldacena1}, much use has been
made of them on both sides, $\mathcal{N}=4$ super Yang-Mills gauge
theory and the $AdS_{5}\times S^{5}$ superstring. The residual symmetry
in light-cone quantization, the centrally extended $\mathfrak{psu}\left(2|2\right)\ltimes\mathbb{R}^{3}$
superalgebra \cite{Beisert2}, has played a very important role in understanding both
sides of the correspondence and the underlying integrability. An important
implication of integrability is that particle momenta are conserved
in scattering, and that every scattering process factorizes into a
sequence of two-particle interactions. Thus all scattering information
is encrypted in the two-body $S$-matrix \cite{Arutyunov2,Staudacher1,Beisert1}.

The requirement that the fundamental $S$-matrix be invariant under
the symmetry algebra fixes it uniquely up to an overall phase factor
\cite{Beisert2} (which must respect unitarity and crossing symmetry).
But in addition to the fundamental particles, the spectrum of the
string sigma model contains an infinite tower of bound states \cite{Dorey2,Dorey1,Arutyunov1}
appearing as poles of the $S$-matrix and in the Bethe ansatz equations.
The construction of $S$-matrices for the bound states is more complicated,
as the symmetry algebra alone is no longer sufficient to determine
the $S$-matrices uniquely. Further constraints are required, arising
from either the Yang-Baxter equation or the underlying Yangian symmetry
\cite{Beisert4,Zwiebel1}.

Yangians are important algebraic structures which appear in many integrable
models \cite{Drinfeld1,Bernard1}, typically as a hidden extension
of Lie symmetry. They are deformations of the polynomial algebra of
the Lie algebra, and are originally associated with integrable models
where the two-particle scattering matrix is a rational function depending
on the difference of rapidities of particles involved in the scattering
\cite{Ragoucy1}. Yangian symmetry has been used to uniquely determine
$S$-matrices describing the scattering of fundamental and bound-state
magnons of closed spin chains \cite{deLeeuw1,Arutyunov6}, and a remnant
of it is expected to govern the scattering from the boundaries as
well. Some charges conserved by this boundary symmetry have recently
been constructed for open spin chains ending on giant graviton
branes with broken symmetries \cite{Ahn1}. It is an interesting challenge
to understand the boundary symmetry in full: an important feature
of quantum integrability is that the presence of suitable boundary
conditions may break a bulk Yangian symmetry without spoiling integrability.
Rather we expect to find a boundary symmetry which is a co-ideal subalgebra
of the bulk symmetry \cite{MacKay1,MacKay2,Doikou1}, probably in
the form of a generalized twisted Yangian \cite{Molev}.

Building on \cite{Ahn1}, our purpose here is to consider the general
framework of boundary Yangian symmetry by considering the reflection
of magnon bound states from the $D3$ brane known as the $Y=0$ maximal
giant graviton \cite{Maldacena2}. Depending on the choice of vacuum
state and relative orientation of the graviton inside $S^{5}$, the
centrally extended $\mathfrak{psu}\left(2|2\right)$ symmetry algebra
may be preserved by the boundary or broken down to $\mathfrak{su}\left(2|1\right)$
\cite{Maldacena2}. In this paper we  construct the hidden boundary
symmetry which extends $\mathfrak{su}\left(2|1\right)$, and find
it to be of the form of the generalized twisted Yangians of \cite{MacKay1,MacKay2}.
The two-magnon bound state reflection matrix which respects this symmetry
proves to be in agreement with \cite{Ahn1}.

We also consider a toy-model boundary which breaks the symmetry down to 
$\mathfrak{su}\left(1|2\right)$. We show that it has a boundary Yangian 
symmetry of the same type as the $Y=0$ giant graviton, but allows 
diagonal reflection only. Thus the reflection matrices for this case
are fully determined by the boundary symmetry alone, and the boundary
Yangian, although a nice mathematical example,  is redundant, in contrast
to earlier case.

This paper is organized as follows. In section 2 we briefly recall
the superspace representation of the symmetry algebra and the Yangian
symmetry of the bulk $S$-matrix. In section 3 we present the general
framework for constructing a generalized twisted boundary Yangian,
and construct the boundary remnant of the bulk Yangian symmetry for
the $Y=0$ giant graviton. Reflection from the boundary
preserving the $\mathfrak{su}\left(1|2\right)$ subalgebra and its corresponding
boundary Yangian and some complementary formulae are presented in the appendices.

\section{Yangian symmetry of the S-matrix}

We begin by briefly reviewing the centrally-extended $\mathfrak{psu}\left(2|2\right)$
algebra and its Yangian extension. This is the symmetry algebra of
the excitations of the light-cone superstring theory on $AdS_{5}\times S^{5}$
(and thereby of the $S$-matrix), and also of the single trace operators
in the $\mathcal{N}=4$ supersymmetric gauge theory that are analogous
to, and known as, spin chains. We shall be using the superspace formalism
introduced in \cite{Arutyunov1}, which simplifies greatly the calculations
of the magnon bound state $S$- and $K$-matrices.

\subsection{Superspace representation of the symmetry algebra}

The centrally-extended $\mathfrak{psu}\left(2|2\right)$ has two sets
of bosonic rotation generators $\mathbb{R}_{a}^{\enskip b}$, $\mathbb{L}_{\alpha}^{\enskip\beta}$,
two sets of fermionic supersymmetry generators $\mathbb{Q}_{\alpha}^{\enskip a},$
$\mathbb{G}_{a}^{\enskip\alpha}$ and three central charges $\mathbb{H}$,
$\mathbb{C}$ and $\mathbb{C}^{\dagger}$. The non-trivial commutation
relations are \cite{Beisert2}
\begin{align}
& \left[\mathbb{L}_{\alpha}^{\enskip\beta},\mathbb{J}_{\gamma}\right]=\delta_{\gamma}^{\beta}\,\mathbb{J}_{\alpha}-\frac{1}{2}\delta_{\alpha}^{\beta}\,\mathbb{J}_{\gamma},
&&\left[\mathbb{L}_{\alpha}^{\enskip\beta},\mathbb{J}^{\gamma}\right]=-\delta_{\alpha}^{\gamma}\,\mathbb{J}^{\beta}+\frac{1}{2}\delta_{\alpha}^{\beta}\,\mathbb{J}^{\gamma},\nonumber \\
& \left[\mathbb{R}_{a}^{\enskip b},\mathbb{J}_{c}\right]=\delta_{c}^{b}\,\mathbb{J}_{a}-\frac{1}{2}\delta_{a}^{b}\,\mathbb{J}_{c},
&&\left[\mathbb{R}_{a}^{\enskip b},\mathbb{J}^{c}\right]=-\delta_{a}^{c}\,\mathbb{J}^{b}+\frac{1}{2}\delta_{a}^{b}\,\mathbb{J}^{c},\nonumber \\
& \left\{ \mathbb{Q}_{\alpha}^{\enskip a},\mathbb{Q}_{\beta}^{\enskip b}\right\} =\epsilon^{ab}\epsilon_{\alpha\beta}\,\mathbb{C},
&&\left\{ \mathbb{G}_{a}^{\enskip\alpha},\mathbb{G}_{b}^{\enskip\beta}\right\} =\epsilon^{\alpha\beta}\epsilon_{ab}\,\mathbb{C}^{\dagger},\nonumber \\
& \left\{ \mathbb{Q}_{\alpha}^{\enskip a},\mathbb{G}_{b}^{\enskip\beta}\right\} =\delta_{b}^{a}\,\mathbb{L}_{\beta}^{\enskip\alpha}
+\delta_{\beta}^{\alpha}\,\mathbb{R}_{b}^{\enskip a}+\delta_{b}^{a}\delta_{\beta}^{\alpha}\,\mathbb{H},\label{psu(2|2)_algebra}
\end{align}
 where $a,\; b,...=1,\;2$ and $\alpha,\;\beta,...=3,\;4$.

A general $l$-magnon bound state is an atypical totally-symmetric
representation of the centrally-extended $\mathfrak{psu}\left(2|2\right)$.
The dimension of the representation is $2l|2l$ and may be neatly
realized as degree $l$ monomial on a graded vector space with the 
basis $\omega_1$, $\omega_2$, $\theta_3$, $\theta_4$,  
where $\omega_{a}$ and $\theta_{\alpha}$ are bosonic and fermionic
variables respectively \cite{Arutyunov1}.

In this representation the centrally-extended $\mathfrak{psu}\left(2|2\right)$
generators are realized as the differential operators
\begin{align}
& \mathbb{R}_{a}^{\enskip b}=\omega_{a}\frac{\partial}{\partial\omega_{b}}-\frac{1}{2}\delta_{a}^{b}\,\omega_{c}\frac{\partial}{\partial\omega_{c}},
&&\mathbb{L}_{\alpha}^{\enskip\beta}=\theta_{\alpha}\frac{\partial}{\partial\theta_{\beta}}-\frac{1}{2}\delta_{\alpha}^{\beta}\,\theta_{\gamma}\frac{\partial}{\partial\theta_{\gamma}},\nonumber \\
& \mathbb{Q}_{\alpha}^{\enskip a}=a\,\theta_{\alpha}\frac{\partial}{\partial\omega_{a}}+b\,\epsilon^{ab}\epsilon_{\alpha\beta}\,\omega_{b}\frac{\partial}{\partial\theta_{\beta}},
&&\mathbb{G}_{a}^{\enskip\alpha}=c\,\epsilon_{ab}\epsilon^{\alpha\beta}\,\theta_{\beta}\frac{\partial}{\partial\omega_{b}}+d\,\omega_{a}\frac{\partial}{\partial\theta_{\alpha}},\nonumber \\
& \mathbb{C}=ab\left(\omega_{a}\frac{\partial}{\partial\omega_{a}}+\theta_{\alpha}\frac{\partial}{\partial\theta_{\alpha}}\right),
&&\mathbb{C}^{\dagger}=cd\left(\omega_{a}\frac{\partial}{\partial\omega_{a}}+\theta_{\alpha}\frac{\partial}{\partial\theta_{\alpha}}\right),\nonumber \\
& \mathbb{H}=\left(ad+bc\right)\left(\omega_{a}\frac{\partial}{\partial\omega_{a}}+\theta_{\alpha}\frac{\partial}{\partial\theta_{\alpha}}\right),
\end{align}
where the coefficients $a=a(p)$, $b=b(p)$, $c=c(p)$, $d=d(p)$
are the representation parameters and the corresponding vector space 
is denoted as $\mathcal{V}^{l}(p,\zeta)$,
where $p$ and $\zeta$ are complex parameters of the representation
and correspond to the momentum and the phase of an individual magnon
in the spin chain.

A convenient parametrization of the representation parameters is%
\footnote{There is a slight abusage of notation here, with $a$, $b$, $c$,
$d$ used both for the representation parameters and bosonic indices,
but these are now the standard conventions. We shall avoid using these
letters anywhere else!%
} \cite{Beisert2}
\begin{equation}
a=\sqrt{\frac{g}{2l}}\eta,\quad b=\sqrt{\frac{g}{2l}}\frac{i\zeta}{\eta}\left(\frac{x^{+}}{x^{-}}-1\right),\quad c=-\sqrt{\frac{g}{2l}}\frac{\eta}{\zeta x^{+}},\quad d=-\sqrt{\frac{g}{2l}}\frac{x^{+}}{i\eta}\left(\frac{x^{-}}{x^{+}}-1\right),\label{abcd}\end{equation}
 where $g$ is a coupling constant, $\zeta={\rm e}^{2i\xi}$ is
the magnon phase and $x^{\pm}$ are the spectral parameters respecting the
mass-shell (multiplet shortening) condition of the $l$-magnon bound state,
\begin{equation}
x^{+}+\frac{1}{x^{+}}-x^{-}-\frac{1}{x^{-}}=i\frac{2l}{g}.\label{shortening}
\end{equation}
 Unitarity requires $\eta={\rm e}^{i\xi}{\rm e}^{i\frac{\varphi}{2}}\sqrt{i\left(x^{-}-x^{+}\right)}$,
where the arbitrary phase factor ${\rm e}^{i\varphi}$ is parameterizing the freedom in choosing $x^{\pm}$.

The eigenvalues of the central charges of $l$-magnon bound state
are expressed as
\begin{eqnarray}
 & C_{l}=l\, ab=\frac{i}{2}g\left({\rm e}^{ip}-1\right){\rm e}^{2i\xi},\qquad C_{l}^{\dagger}=l\, cd=-\frac{i}{2}g\left({\rm e}^{-ip}-1\right){\rm e}^{-2i\xi},\nonumber \\
 & H_{l}=l\left(ad+bc\right)=\sqrt{l^{2}+4g^{2}\sin^{2}\frac{p}{2}}.
\end{eqnarray}

We write the $S$-matrix as a differential operator in superspace
acting on the tensor product of two vector spaces
\begin{equation}
S(p_{1},p_{2}):\quad\mathcal{V}^{M}(p_{1},\zeta)\otimes\mathcal{V}^{N}(p_{2},\zeta{\rm e}^{ip_{1}})\rightarrow
\mathcal{V}^{M}(p_{1},\zeta{\rm e}^{ip_{2}})\otimes\mathcal{V}^{N}(p_{2},\zeta),\label{S-matrix}
\end{equation}
where we have chosen phase $\zeta$ to increase from left to right.
In the superspace formalism the $S$-matrix may be viewed as an element
of
\begin{equation}
End\left(\mathcal{V}^{M}\otimes\mathcal{V}^{N}\right)\approx\mathcal{V}^{M}\otimes\mathcal{V}^{N}\otimes\mathcal{D}_{M}\otimes\mathcal{D}_{N},
\end{equation}
 where $\mathcal{D}_{M}$ is the vector space dual to $\mathcal{V}^{M}$.
The dual vector space is realized as the space of polynomials of degree
$M$ of the differential operators $\frac{\partial}{\partial\omega_{a}}$
and $\frac{\partial}{\partial\theta_{\alpha}}$, with a natural pairing
between $\mathcal{D}_{M}$ and $\mathcal{V}^{M}$ induced by the relations
$\frac{\partial}{\partial\omega_{a}}\omega_{b}=\delta_{b}^{a}$ and
$\frac{\partial}{\partial\theta_{\alpha}}\theta_{\beta}=\delta_{\beta}^{\alpha}$. 
Thus the $S$-matrix may be represented as
\begin{equation}
S(p_{1},p_{2})=\sum_{i}a_{i}(p_{1},p_{2})\,\Lambda_{i},
\end{equation}
 where $\Lambda_{i}$ span a complete basis of differential operators
invariant under the $\mathfrak{su}\left(2\right)\oplus\mathfrak{su}\left(2\right)$
algebra and $a_{i}(p_{1},p_{2})$ are $S$-matrix components.
The exact expressions of $\Lambda_{i}$ for various $S$-matrices
and general formulas how to compute generic $\Lambda_{i}$ are given
in \cite{Arutyunov1}.

The invariance of the S-matrix under the co-products 
of the generators of the symmetry algebra reads as
\begin{equation}
S(p_{1},p_{2})\,\Delta\!\left(\mathbb{J}^{A}\right)=\Delta^{op}\!\left(\mathbb{J}^{A}\right)S(p_{1},p_{2}),\label{S_invariance}
\end{equation}
 where $\Delta^{op}=P\Delta P$ with $P$ being a graded permutation.
The invariance constrains all coefficients of the $S$-matrix of the fundamental
states up to an overall phase. In the case of scattering of $l$-
with $m$-bound states with $l,\; m\geq2$, the symmetry algebra alone
is not enough to fix all $S$-matrix coefficients, and additional
constraints are required from the Yang-Baxter equation or, alternatively, Yangian symmetry
\cite{Arutyunov1,Arutyunov6}.

\subsection{Yangian symmetry and co-products}

The Yangian $\mbox{Y}(\mathfrak{g})$ of a Lie algebra
$\mathfrak{g}$ is a deformation of the universal enveloping algebra
$\mbox{U}\left(\mathfrak{g}[u]\right)$ of the polynomial
algebra $\mathfrak{g}[u]$. It is generated by grade-0
$\mathfrak{g}$ generators $\mathbb{J}^{A}$ and grade-1 $\mbox{Y}(\mathfrak{g})$
generators $\hat{\mathbb{J}}^{A}$. Their commutators have the generic
form
\begin{equation}
\left[\mathbb{J}^{A},\mathbb{J}^{B}\right]=f_{\quad C}^{AB}\,\mathbb{J}^{C},
\qquad\bigl[\mathbb{J}^{A},\hat{\mathbb{J}}^{B}\bigr]=f_{\quad C}^{AB}\,\hat{\mathbb{J}}^{C},
\end{equation}
 and must obey Jacobi and Serre relations
\begin{align}
\bigl[\mathbb{J}^{[A},\bigl[\mathbb{J}^{B},\mathbb{J}^{C]}\bigr]\bigr] & =0,
\qquad\bigl[\mathbb{J}^{[A},\bigl[\mathbb{J}^{B},\hat{\mathbb{J}}^{C]}\bigr]\bigr]=0,\nonumber \\
\bigl[\hat{\mathbb{J}}^{[A},\bigl[\hat{\mathbb{J}}^{B},\mathbb{J}^{C]}\bigr]\bigr] 
& =\frac{1}{4}f_{\quad D}^{AG}f_{\quad E}^{BH}f_{\quad F}^{CK}f_{GHK}\,\mathbb{J}^{\{D}\mathbb{J}^{E}\mathbb{J}^{F\}}.
\end{align}
 The indices of structure constants $f_{\quad D}^{AB}$ are lowered
by the means of the inverse Killing-Cartan form $g_{BD}$. In the
case of interest $\mathfrak{g}$ is the centrally-extended $\mathfrak{psu}\left(2|2\right)$,
and the relevant Killing form is degenerate. However, this degeneracy
may be cured in several ways---for example, by considering the limit $\varepsilon\rightarrow0$ 
of the exceptional superalgebra $\mathfrak{d}(2,1;\varepsilon)$ \cite{Matsumoto1}, 
or by considering Drinfeld's second realization of the centrally-extended $\mathfrak{psu}\left(2|2\right)$,
which was shown to be isomorphic to the first realization in \cite{Spill1}.

The co-products of the grade-0 and grade-1 generators take form
\begin{align}
\Delta\mathbb{J}^{A} & =\mathbb{J}^{A}\otimes1+1\otimes\mathbb{J}^{A},\qquad
\Delta\hat{\mathbb{J}}^{A}=\hat{\mathbb{J}}^{A}\otimes1+1\otimes\hat{\mathbb{J}}^{A}+\frac{1}{2}f_{\; BC}^{A}\,\mathbb{J}^{B}\otimes\mathbb{J}^{C}.
\end{align}

Crucial in constructing the finite-dimensional representations of $\mbox{Y}(\mathfrak{g})$ is the one-para\-meter family of the `evaluation automorphisms' \begin{equation}
\tau_v: \mbox{Y}(\mathfrak{g})\rightarrow\mbox{Y}(\mathfrak{g})\quad \mathbb{J}^{A}\mapsto\mathbb{J}^{A}\,,\quad \hat{\mathbb{J}}^{A}\mapsto \hat{\mathbb{J}}^{A}+v \mathbb{J}^{A}\,,
\end{equation}
corresponding to a shift in the polynomial variable, which implies that $\mbox{Y}(\mathfrak{g})$ representations appear in one-parameter families. On (the limited set of) finite--dimensional irreducible
representations of $\mathfrak{g}$ which may be extended to representations of $\mbox{Y}(\mathfrak{g})$, these
are realized via the `evaluation map' \begin{equation}\mathrm{ev}_v:\mbox{Y}(\mathfrak{g})\mapsto\mbox{U}\left(\mathfrak{g}\right)\quad \mathbb{J}^{A}\mapsto\mathbb{J}^{A}\,,\quad \hat{\mathbb{J}}^{A}\mapsto v \mathbb{J}^{A}\,,
\end{equation}
which yields `evaluation modules', with states $\left|v\right\rangle $ carrying
a spectral parameter $v$.
As was shown in \cite{Beisert4}, the magnon states are of this form, and have
\begin{equation}
\hat{\mathbb{J}}^{A}\left|u\right\rangle =i\frac{g}{2}u\,\mathbb{J}^{A}\left|u\right\rangle ,\label{uJ_ansatz}
\end{equation}
 where $u$ is the rapidity of the corresponding magnon state. In
the case of the bulk $l$-magnon bound states of the centrally-extended
$\mathfrak{psu}\left(2|2\right)$, the rapidity is $u\equiv u(p)=x^{+}+\frac{1}{x^{+}}-i\frac{l}{g}$.

The Yangian symmetry fixes the bound state $S$-matrices uniquely up to
an overall phase by requiring their invariance under the co-products
of the Yangian generators \cite{Beisert4}
\begin{align}
\Delta\hat{\mathbb{R}}_{a}^{\enskip b} & =\hat{\mathbb{R}}_{a}^{\enskip b}\otimes1+1\otimes\hat{\mathbb{R}}_{a}^{\enskip b}+\frac{1}{2}\mathbb{R}_{a}^{\enskip c}\otimes\mathbb{R}_{c}^{\enskip b}-\frac{1}{2}\mathbb{R}_{c}^{\enskip b}\otimes\mathbb{R}_{a}^{\enskip c}-\frac{1}{2}\mathbb{G}_{a}^{\enskip\gamma}\otimes\mathbb{Q}_{\gamma}^{\enskip b}-\frac{1}{2}\mathbb{Q}_{\gamma}^{\enskip b}\otimes\mathbb{G}_{a}^{\enskip\gamma}\nonumber \\
 & \qquad+\frac{1}{4}\delta_{a}^{b}\mathbb{G}_{c}^{\enskip\gamma}\otimes\mathbb{Q}_{\gamma}^{\enskip c}+\frac{1}{4}\delta_{a}^{b}\mathbb{Q}_{\gamma}^{\enskip c}\otimes\mathbb{G}_{c}^{\enskip\gamma},\nonumber \\
\Delta\hat{\mathbb{L}}_{\alpha}^{\enskip\beta} & =\hat{\mathbb{L}}_{\alpha}^{\enskip\beta}\otimes1+1\otimes\hat{\mathbb{L}}_{\alpha}^{\enskip\beta}-\frac{1}{2}\mathbb{L}_{\alpha}^{\enskip\gamma}\otimes\mathbb{L}_{\gamma}^{\enskip\beta}+\frac{1}{2}\mathbb{L}_{\gamma}^{\enskip\beta}\otimes\mathbb{L}_{\alpha}^{\enskip\gamma}+\frac{1}{2}\mathbb{G}_{c}^{\enskip\beta}\otimes\mathbb{Q}_{\alpha}^{\enskip c}+\frac{1}{2}\mathbb{Q}_{\alpha}^{\enskip c}\otimes\mathbb{G}_{c}^{\enskip\beta}\nonumber \\
 & \qquad-\frac{1}{4}\delta_{\alpha}^{\beta}\mathbb{G}_{c}^{\enskip\gamma}\otimes\mathbb{Q}_{\gamma}^{\enskip c}-\frac{1}{4}\delta_{\alpha}^{\beta}\mathbb{Q}_{\gamma}^{\enskip c}\otimes\mathbb{G}_{c}^{\enskip\gamma},\nonumber \\
\Delta\hat{\mathbb{Q}}_{\alpha}^{\enskip a} & =\hat{\mathbb{Q}}_{\alpha}^{\enskip a}\otimes1+1\otimes\hat{\mathbb{Q}}_{\alpha}^{\enskip a}+\frac{1}{2}\mathbb{Q}_{\alpha}^{\enskip c}\otimes\mathbb{R}_{c}^{\enskip a}-\frac{1}{2}\mathbb{R}_{c}^{\enskip a}\otimes\mathbb{Q}_{\alpha}^{\enskip c}+\frac{1}{2}\mathbb{Q}_{\gamma}^{\enskip a}\otimes\mathbb{L}_{\alpha}^{\enskip\gamma}-\frac{1}{2}\mathbb{L}_{\alpha}^{\enskip\gamma}\otimes\mathbb{Q}_{\gamma}^{\enskip a}\nonumber \\
 & \qquad+\frac{1}{4}\mathbb{Q}_{\alpha}^{\enskip a}\otimes\mathbb{H}-\frac{1}{4}\mathbb{H}\otimes\mathbb{Q}_{\alpha}^{\enskip a}+\frac{1}{2}\varepsilon_{\alpha\gamma}\varepsilon^{ac}\mathbb{C}\otimes\mathbb{G}_{c}^{\enskip\gamma}-\frac{1}{2}\varepsilon_{\alpha\gamma}\varepsilon^{ac}\mathbb{G}_{c}^{\enskip\gamma}\otimes\mathbb{C},\nonumber \\
\Delta\hat{\mathbb{G}}_{a}^{\enskip\alpha} & =\hat{\mathbb{G}}_{a}^{\enskip\alpha}\otimes1+1\otimes\hat{\mathbb{G}}_{a}^{\enskip\alpha}-\frac{1}{2}\mathbb{G}_{c}^{\enskip\alpha}\otimes\mathbb{R}_{a}^{\enskip c}+\frac{1}{2}\mathbb{R}_{a}^{\enskip c}\otimes\mathbb{G}_{c}^{\enskip\alpha}-\frac{1}{2}\mathbb{G}_{a}^{\enskip\gamma}\otimes\mathbb{L}_{\gamma}^{\enskip\alpha}+\frac{1}{2}\mathbb{L}_{\gamma}^{\enskip\alpha}\otimes\mathbb{G}_{a}^{\enskip\gamma}\nonumber \\
 & \qquad-\frac{1}{4}\mathbb{G}_{a}^{\enskip\alpha}\otimes\mathbb{H}+\frac{1}{4}\mathbb{H}\otimes\mathbb{G}_{a}^{\enskip\alpha}-\frac{1}{2}\varepsilon_{ac}\varepsilon^{\alpha\gamma}\mathbb{C}^{\dagger}\otimes\mathbb{Q}_{\gamma}^{\enskip c}+\frac{1}{2}\varepsilon_{ac}\varepsilon^{\alpha\gamma}\mathbb{Q}_{\gamma}^{\enskip c}\otimes\mathbb{C}^{\dagger},\nonumber \\
\Delta\hat{\mathbb{C}} & =\hat{\mathbb{C}}\otimes1+1\otimes\hat{\mathbb{C}}-\frac{1}{2}\mathbb{H}\otimes\mathbb{C}+\frac{1}{2}\mathbb{C}\otimes\mathbb{H},\nonumber \\
\Delta\hat{\mathbb{C}}^{\dagger} & =\hat{\mathbb{C}}^{\dagger}\otimes1+1\otimes\mathbb{\hat{C}}^{\dagger}+\frac{1}{2}\mathbb{H}\otimes\mathbb{C}^{\dagger}-\frac{1}{2}\mathbb{C}^{\dagger}\otimes\mathbb{H},\nonumber \\
\Delta\hat{\mathbb{H}} & =\hat{\mathbb{H}}\otimes1+1\otimes\hat{\mathbb{H}}+\mathbb{C}\otimes\mathbb{C}^{\dagger}-\mathbb{C}^{\dagger}\otimes\mathbb{C}.\label{Y(g)}
\end{align}
 The non-trivial braiding factors are not explicitly shown in the
co-products above---rather they are all hidden in the parameters of
the representation (\ref{abcd}) and the choice of phase in (\ref{S-matrix}).

\section{Boundary remnants of Yangian symmetry}

In this section we shall present a method for constructing a boundary
remnant of bulk Yangian symmetry, which takes the form of a generalized
twisted Yangian $\mbox{Y}(\mathfrak{g},\mathfrak{h})$,
constructed as a subalgebra of $\mbox{Y}(\mathfrak{g})$.
Here $\mathfrak{h\subset g}$ is the Lie subalgebra preserved by
the boundary of the Lie algebra $\mathfrak{g}$ respected by the
bulk states \cite{MacKay1,MacKay2}. The construction requires that
 $\left(\mathfrak{g},\mathfrak{h}\right)$
be a symmetric pair (\ref{symmetric_pair}) \cite{MacKay3}. The centrally-extended
algebra $\mathfrak{g=psu}\left(2|2\right)\ltimes\mathbb{R}^{3}$ of
the light-cone string may be split into a symmetric pair in two algebraically
similar ways, with $\mathfrak{h}=\mathfrak{su}\left(2|1\right)$ and
$\mathfrak{h}=\mathfrak{su}\left(1|2\right)$, but they are very different
from the scattering theory point
of view.

The first case, $\mathfrak{h}=\mathfrak{su}\left(2|1\right)$, corresponds
to an open string ending on the $Y=0$ giant graviton and leads to
the reflection matrix governed by the boundary Yangian symmetry. We
shall consider this case in the subsection 3.2.

The second case, $\mathfrak{h}=\mathfrak{su}\left(1|2\right)$,
which we consider as a toy model, leads
to a trivial (diagonal) reflection matrix fully determined by the
symmetry algebra alone. Although it still possesses a boundary Yangian
symmetry, this symmetry now appears redundant. We present it in
the appendix A as a mathematical exercise; its role in the AdS/CFT
correspondence is not clear.

\subsection{Boundary Yangian symmetry}

Consider an integrable boundary field theory which preserves only
a subalgebra $\mathfrak{h\subset g}$ of the symmetry respected by
the bulk fields. We want to find the corresponding Yangian
charges conserved under the reflection. For this purpose, we shall
be considering the boundary Yangian algebra acting on the evaluation
module (\ref{uJ_ansatz}) carrying the spectral parameter $u$ which
is mapped to the rapidity of the state in field theory. The reflection
in the integrable field theory results in a change of sign of the
rapidity $u\mapsto-u$.

Integrability requires that the boundary Lie symmetry $\mathfrak{h}\subset\mathfrak{g}$ be a subalgebra invariant under an involution $\sigma$ \cite{MacKay3}. 
Thus we proceed by splitting the bulk algebra into $\mathfrak{g}=\mathfrak{h}\oplus\mathfrak{m}$
under a graded involution $\sigma$ of $\mathfrak{g}$ with the
eigenspaces $\sigma(\mathfrak{h})=+1$ and $\sigma(\mathfrak{m})=-1$.
One then has
\begin{equation}
\left[\mathfrak{h},\mathfrak{h}\right]\subset\mathfrak{h},
\qquad\left[\mathfrak{h},\mathfrak{m}\right]\subset\mathfrak{m},
\qquad\left[\mathfrak{m},\mathfrak{m}\right]\subset\mathfrak{h},\label{symmetric_pair}
\end{equation}
together with orthogonality with respect to the Killing form, $\kappa(\mathfrak{h},\mathfrak{m})=0$.
This is crucial in guaranteeing the co-ideal property -- that the co-product
of any Yangian charge $\hat{\mathbb{J}}$ preserved by the boundary
must be in the tensor product of bulk and boundary Yangian
\begin{equation}
\Delta\hat{\mathbb{J}}\in\mbox{Y}(\mathfrak{g})\otimes\mbox{Y}(\mathfrak{g},\mathfrak{h}).\label{coideal}
\end{equation}
This ensures that multiparticle products of bulk and boundary states
still represent $\mbox{Y}(\mathfrak{g},\mathfrak{h})$,
and is analogous to the requirement that $\Delta$ be a homomorphism
$\mbox{Y}(\mathfrak{g})\rightarrow\mbox{Y}(\mathfrak{g})\otimes\mbox{Y}(\mathfrak{g})$
imposed by multiparticle bulk states.

Just as $\mbox{Y}(\mathfrak{g})$ was a deformation of
$\mbox{U}\left(\mathfrak{g}[u]\right)$, so $\mbox{Y}(\mathfrak{g},\mathfrak{h})$
may be thought of as a deformation of the subalgebra of $\mbox{U}\left(\mathfrak{g}[u]\right)$
which is invariant under the extension $\bar{\sigma}$ of $\sigma$
which sends $\bar{\sigma}:u\mapsto-u$ (representing the change of
sign of the magnon rapidity after the reflection),
\begin{equation}
\mathfrak{h}\oplus u\mathfrak{m}\oplus...\subset\mathfrak{g}[u]=\left(\mathfrak{h}\oplus\mathfrak{m}\right)\oplus u\left(\mathfrak{h}\oplus\mathfrak{m}\right)\oplus...
\end{equation}
Hence the boundary Yangian charges must live in the subspace $u\mathfrak{m}$.
However, while the grade-0 generators of $\mathfrak{h}$ clearly respect
the co-ideal property, the grade-1 generators of $u\mathfrak{m}$ do
not do so,
\begin{equation}
\Delta\hat{\mathbb{J}}^{p}=\hat{\mathbb{J}}^{p}\otimes1+1\otimes\hat{\mathbb{J}}^{p}+\frac{1}{2}f_{\; qi}^{p}\left(\mathbb{J}^{q}\otimes\mathbb{J}^{i}+\mathbb{J}^{i}\otimes\mathbb{J}^{q}\right)\notin\mbox{Y}(\mathfrak{g})\otimes\mbox{Y}(\mathfrak{g},\mathfrak{h}),
\end{equation}
where $i(,j,k,...)$ run over the $\mathfrak{h}$-indices and $p,q(,r,...)$
over the $\mathfrak{m}$-indices. Rather we need a deformation of
the grade-1 $u\mathfrak{m}$ generators, and therefore we find $\mbox{Y}(\mathfrak{g},\mathfrak{h})$
to be the algebra generated by $\{\mathbb{J}^{i},\,\tilde{\mathbb{J}}^{p}\}$,
where
\begin{equation}
\tilde{\mathbb{J}}^{p}:=\hat{\mathbb{J}}^{p}+\frac{1}{2}f_{\; qi}^{p}\,\mathbb{J}^{q}\,\mathbb{J}^{i},\label{twist}
\end{equation}
are the twisted boundary Yangian generators. (One can arrive at the same deformation in different ways --- for example, via the boundary transfer matrix, or by considering the charges' classical conservation \cite{MacKay2}.)

Now we can show that $\mbox{Y}(\mathfrak{g},\mathfrak{h})$
is a left co-ideal subalgebra, $\Delta\mbox{Y}(\mathfrak{g},\mathfrak{h})\subset\mbox{Y}(\mathfrak{g})\otimes\mbox{Y}(\mathfrak{g},\mathfrak{h})$.
To do this one calculates explicitly the co-product of the twisted
Yangian generator,
\begin{eqnarray}
\Delta\tilde{\mathbb{J}}^{p} & = & \Delta\hat{\mathbb{J}}^{p}+\frac{1}{2}f_{\; qi}^{p}\,\Delta\mathbb{J}^{q}\Delta\mathbb{J}^{i}\nonumber \\
 & = & \hat{\mathbb{J}}^{p}\otimes1+1\otimes\hat{\mathbb{J}}^{p}+\frac{1}{2}f_{\; qi}^{p}\left(\mathbb{J}^{q}\,\mathbb{J}^{i}\otimes1+1\otimes\mathbb{J}^{q}\,\mathbb{J}^{i}\right)\nonumber \\
 &  & +\frac{1}{2}f_{\; iq}^{p}\,\mathbb{J}^{i}\otimes\mathbb{J}^{q}+\frac{1}{2}f_{\; qi}^{p}\,\mathbb{J}^{q}\otimes\mathbb{J}^{i}+\frac{1}{2}f_{\; qi}^{p}\left(\mathbb{J}^{q}\otimes\mathbb{J}^{i}+\mathbb{J}^{i}\otimes\mathbb{J}^{q}\right)\nonumber \\
 & = & \tilde{\mathbb{J}}^{p}\otimes1+1\otimes\tilde{\mathbb{J}}^{p}+f_{\; qi}^{p}\,\mathbb{J}^{q}\otimes\mathbb{J}^{i}\nonumber \\
 & \in & \mbox{Y}(\mathfrak{g})\otimes\mbox{Y}(\mathfrak{g},\mathfrak{h}),\label{m-coideal}
\end{eqnarray}
 where we have used the implication of the symmetric pair decomposition
(\ref{symmetric_pair}) that the only non-zero structure constants
are $f^{pqi}$ and $f^{pij}$.

We should emphasize that the map $\bar{\sigma}$ together with the
twist (\ref{twist}) is an automorphism of $\mathfrak{g}[u]$
realized on the `evaluation module' (\ref{uJ_ansatz}) and is independent
of its explicit realization in field theory, {\em i.e.\ }it is \textit{not} a
map on the fields.

\subsection{Yangian symmetry of the Y=0 maximal giant graviton}

Maximal giant gravitons are $D3$ branes in $AdS_{5}\times S^{5}$
which wraps a topologically-trivial cycles comprising a maximal $S^{3}$
within the $S^{5}$. Giant gravitons are prevented from collapsing
by their coupling to the background supergravity fields. The usual
parametrization of the $S^{5}$ is expressed in terms of $X=\Phi_{1}+i\Phi_{2}$,
$Y=\Phi_{3}+i\Phi_{4}$, $Z=\Phi_{5}+i\Phi_{6}$ respecting $|X|^{2}+|Y|^{2}+|Z|^{2}=1$,
where the radius of the $S^{5}$ has been normalized to $R=1$. In
this parametrization the maximal giant graviton may be obtained by
setting any two $\Phi_{i}$ to zero.

Any two such configurations are of course related by an $SO\left(6\right)$
rotation. However, one can break this equivalence by attaching an
open string to the brane and giving the string a charge $J$ corresponding
to a preferred $SO\left(2\right)\subset SO\left(6\right)$ rotation.
In the limit when $J$ is large the field theory description of the
string carries a large number of insertions of the field corresponding
to the preferred rotation. It was shown in \cite{Maldacena2} that
the explicit description of the open string depends on the selection
of a particular generator $J$ and the relevant orientation of the
giant graviton inside $S^{5}$. The two interesting cases are given
by choosing the charge to be $J=J_{56}$ and the giant graviton to
be a three sphere given by $Y=0$ or $Z=0$.

The $Y=0$ giant graviton preserves the subgroup which is also preserved
by the field $Y$. This restricts the symmetry algebra on the boundary
to be $\mathfrak{h=su}\left(2|1\right)$ and has no degrees of freedom
attached to the end of the spin chain \cite{Maldacena2}. The commutation
relations of $\mathfrak{su}\left(2|1\right)$ are acquired from (\ref{psu(2|2)_algebra})
by dropping the generators with bosonic indices $a,\, b,\, c,\;...=2$;
thus the surviving generators are $\mathbb{L}_{\alpha}^{\enskip\beta},$
$\mathbb{R}_{1}^{\enskip1},$ $\mathbb{R}_{2}^{\enskip2}\equiv-\mathbb{R}_{1}^{\enskip1},$
$\mathbb{Q}_{\alpha}^{\enskip1}$, $\mathbb{G}_{1}^{\enskip\alpha}$
and $\mathbb{H}$. It is straightforward to check that the subalgebra
$\mathfrak{h=su}\left(2|1\right)$ and subset $\mathfrak{m}=\mathfrak{psu}\left(2|2\right)\ltimes\mathbb{R}^{3}/\mathfrak{su}\left(2|1\right)$
consisting of generators $\mathbb{R}_{1}^{\enskip2},\;\mathbb{R}_{2}^{\enskip1},\;\mathbb{Q}_{\gamma}^{\enskip2},\;\mathbb{G}_{2}^{\enskip\gamma},\;\mathbb{C},\;\mathbb{C}^{\dagger}$
form a symmetric pair (\ref{symmetric_pair}) by considering the commutation
relations (\ref{psu(2|2)_algebra}). Hence the theory should possess
a boundary Yangian.

We shall construct the scattering theory first. Following \cite{Ahn1}
we define a boundary vacuum state $\left|0_{B}\right\rangle$
and a corresponding trivial vector space $\mathcal{V}_{B}(0)$
which is annihilated by all $\mathfrak{su}\left(2|1\right)$ generators.
This allows us to define the superspace $K$-matrix for the reflection
of bulk magnons from the boundary vacuum state as an operator acting
on the tensor product of spaces,
\begin{equation}
K(p):\quad\mathcal{V}^{M}(p,\zeta)\otimes\mathcal{V}_{B}(0)\rightarrow\mathcal{V}^{M}(-p,\zeta)\otimes\mathcal{V}_{B}(0),\label{K_braiding}
\end{equation}
 where the reflection matrix is defined as a differential operator
\begin{equation}
K(p)=\sum_{i}k_{i}(p)\,\Lambda_{i}
\end{equation}
 acting on the superspace. In the case of the reflection of fundamental
states, the symmetry algebra implies that the the only dynamics allowed is 
$\omega_{i}\rightarrow\omega_{i}$ and $\theta_{\alpha}\rightarrow\theta_{\alpha}$ 
with different reflection coefficients 
for $\omega_{1}$ and $\omega_{2}$, but the same for $\theta_{3}$ and $\theta_{4}$.
Hence the reflection matrix $K^{A}$ may be represented on a superspace as
\begin{equation}
K^{A}(p)=k_{1}(p)\,\omega_{1}\frac{\partial}{\partial\omega_{1}}+k_{2}(p)\,\omega_{2}\frac{\partial}{\partial\omega_{2}}+k_{3}(p)\,\theta_{\alpha}\frac{\partial}{\partial\theta_{\alpha}}.\end{equation}
The boundary symmetry algebra fixes the reflection coefficients
(up to an overall factor) to be
\begin{equation}
k_{1}(p)=1,\qquad k_{2}(p)=-\frac{x^{-}}{x^{+}},\qquad k_{3}(p)=\frac{\tilde{\eta}}{\eta}.
\end{equation}
 In the case of the reflection of two-magnon bound states, reflection
matrix $K^{B}$ is no longer diagonal. The diagonal reflection channels are 
$ \omega_{i}\omega_{j}\rightarrow\omega_{i}\omega_{j}$, $\omega_{i}\theta_{\alpha}\rightarrow\omega_{i}\theta_{\alpha}$,
$\theta_{3}\theta_{4}\rightarrow\theta_{3}\theta_{4}$, and the off-diagonal 
are $\omega_{1}\omega_{2}\rightarrow\theta_{3}\theta_{4}$ and $\theta_{3}\theta_{4}\rightarrow\omega_{1}\omega_{2}$.
This leads to the following representation of the reflection matrix on the superspace
\begin{equation}
K^{B}(p)=\sum_{i=1}^{8}k_{i}(p)\,\Lambda_{i},
\end{equation}
 where $\Lambda_{i}$ with $i=1,...\,,6$ are diagonal and $\Lambda_{7}$,
$\Lambda_{8}$ are off-diagonal differential operators
\begin{align}
& \Lambda_{1}=\frac{1}{2}\omega_{1}\omega_{1}\frac{\partial^{2}}{\partial\omega_{1}\partial\omega_{1}},
&& \Lambda_{2}=\omega_{1}\omega_{2}\frac{\partial^{2}}{\partial\omega_{2}\partial\omega_{1}},
&& \Lambda_{3}=\frac{1}{2}\omega_{2}\omega_{2}\frac{\partial^{2}}{\partial\omega_{2}\partial\omega_{2}},\nonumber \\
& \Lambda_{4}=\theta_{3}\theta_{4}\frac{\partial^{2}}{\partial\theta_{4}\partial\theta_{3}},
&& \Lambda_{5}=\omega_{1}\theta_{\alpha}\frac{\partial^{2}}{\partial\omega_{1}\partial\theta_{\alpha}},
&& \Lambda_{6}=\omega_{2}\theta_{\alpha}\frac{\partial^{2}}{\partial\omega_{2}\partial\theta_{\alpha}}.\nonumber \\
& \Lambda_{7}=\theta_{3}\theta_{4}\frac{\partial^{2}}{\partial\omega_{2}\partial\omega_{1}},
&& \Lambda_{8}=\omega_{1}\omega_{2}\frac{\partial^{2}}{\partial\theta_{4}\partial\theta_{3}}.
\end{align}
 In this case, following the general pattern \cite{Arutyunov1,Arutyunov6},
the symmetry algebra alone is not enough to fix all reflection coefficients
uniquely. This is the consequence of the relation between representations
of $\mathfrak{su}\left(2|2\right)$ and $\mathfrak{su}\left(2|1\right)$.
Fundamental magnons  transform irreducibly in the fundamental
representation $\boxslash$ of $\mathfrak{su}\left(2|2\right)$ and
in the supersymmetric representation $\boxslash\!\boxslash$ of $\mathfrak{su}\left(2|1\right)$.
This is no longer the case for the two-magnon bound states, which
transform irreducibly in a supersymmetric representation $\boxslash\!\boxslash$
of $\mathfrak{su}\left(2|2\right)$, but in a {\em reducible}
representation ${\boxslash}\!{\boxslash}\!{\boxslash}\!{\boxslash}\, {\oplus} {\renewcommand{\arraystretch}{0}\begin{array}{c}
{\boxslash\!\boxslash}\\
{\boxslash\!\boxslash}\end{array}}$ of $\mathfrak{su}\left(2|1\right)$: and one further needs either
the boundary Yang-Baxter equation or boundary Yangian symmetry to
fix the ratio between the representations of $\mathfrak{su}\left(2|1\right)$.
The number of reducible components grows with bound state number
$l$. In the case of reflection of two-magnon bound states the symmetry
algebra fixes 7 out of 8 reflection coefficients up to an overall dressing
phase. Hence one needs to impose only one additional constraint to
fix the last coefficient. A conserved Yangian charge $\tilde{\mathbb{Q}}$
giving the required constraint was constructed and the reflection
matrix $K^{B}$ was calculated in \cite{Ahn1} --- but one charge alone would not typically
be enough to constrain uniquely the higher-order bound state $S$-matrices.

We shall construct the boundary Yangian $\mbox{Y}(\mathfrak{g},\mathfrak{h})$
using  (\ref{twist}). The co-product of twisted
Yangian generators $\Delta\tilde{\mathbb{J}}^{p}$ acts on the tensor
product of bulk and boundary vector spaces of bulk and boundary algebra
and, by the construction above, any generator of the boundary
symmetry annihilates the boundary vacuum state $\left|0_{B}\right\rangle $.
Hence the non-trivial parts of the co-products of twisted Yangian generators are
\begin{align}
\Delta\tilde{\mathbb{R}}_{1}^{\enskip2} & =\left(\hat{\mathbb{R}}_{1}^{\enskip2}+\frac{1}{2}\mathbb{R}_{1}^{\enskip2}\,\mathbb{R}_{1}^{\enskip1}-\frac{1}{2}\mathbb{R}_{1}^{\enskip2}\,\mathbb{R}_{2}^{\enskip2}-\frac{1}{2}\mathbb{Q}_{\gamma}^{\enskip2}\,\mathbb{G}_{1}^{\enskip\gamma}\right)\otimes1,\nonumber \\
\Delta\tilde{\mathbb{R}}_{2}^{\enskip1} & =\left(\hat{\mathbb{R}}_{2}^{\enskip1}+\frac{1}{2}\mathbb{R}_{2}^{\enskip1}\,\mathbb{R}_{1}^{\enskip1}-\frac{1}{2}\mathbb{R}_{2}^{\enskip1}\,\mathbb{R}_{2}^{\enskip2}-\frac{1}{2}\mathbb{G}_{2}^{\enskip\gamma}\,\mathbb{Q}_{\gamma}^{\enskip1}\right)\otimes1,\nonumber \\
\Delta\tilde{\mathbb{Q}}_{\alpha}^{\enskip2} & =\left(\hat{\mathbb{Q}}_{\alpha}^{\enskip2}+\frac{1}{2}\mathbb{Q}_{\alpha}^{\enskip2}\,\mathbb{R}_{2}^{\enskip2}-\frac{1}{2}\mathbb{R}_{1}^{\enskip2}\,\mathbb{Q}_{\alpha}^{\enskip1}+\frac{1}{2}\mathbb{Q}_{\gamma}^{\enskip2}\,\mathbb{L}_{\alpha}^{\enskip\gamma}+\frac{1}{4}\mathbb{Q}_{\alpha}^{\enskip2}\,\mathbb{H}-\frac{1}{2}\varepsilon_{\alpha\gamma}\mathbb{C}\,\mathbb{G}_{1}^{\enskip\gamma}\right)\otimes1,\nonumber \\
\Delta\tilde{\mathbb{G}}_{2}^{\enskip\alpha} & =\left(\hat{\mathbb{G}}_{2}^{\enskip\alpha}-\frac{1}{2}\mathbb{G}_{2}^{\enskip\alpha}\,\mathbb{R}_{2}^{\enskip2}+\frac{1}{2}\mathbb{R}_{2}^{\enskip1}\,\mathbb{G}_{1}^{\enskip\alpha}-\frac{1}{2}\mathbb{G}_{2}^{\enskip\gamma}\,\mathbb{L}_{\gamma}^{\enskip\alpha}-\frac{1}{4}\mathbb{G}_{2}^{\enskip\alpha}\,\mathbb{H}+\frac{1}{2}\varepsilon^{\alpha\gamma}\mathbb{C}^{\dagger}\,\mathbb{Q}_{\gamma}^{\enskip1}\right)\otimes1,\nonumber \\
\Delta\tilde{\mathbb{C}} & =\left(\hat{\mathbb{C}}+\frac{1}{2}\mathbb{C}\,\mathbb{H}\right)\otimes1,\nonumber \\
\Delta\mathbb{\tilde{C}}^{\dagger} & =\left(\hat{\mathbb{C}}^{\dagger}-\frac{1}{2}\mathbb{C}^{\dagger}\,\mathbb{H}\right)\otimes1.\label{Y(g,h)}
\end{align}
The braiding factors are defined by the reflection equation (\ref{K_braiding})\footnote{The braiding
for the left factor of the co-products is always trivial because the reflection results only
in a change of sign for momentum $p\mapsto {-}p$ of the incoming magnon, {\em i.e.\ }maps spectral parameters
$x^{\pm}\mapsto {-}x^{\mp}$. The braiding would be non--trivial in the right factor of 
co-products corresponding to boundaries with degrees of freedom.}
and are hidden in the representation parameters (\ref{abcd}).
The full expressions of the co-products is given in the appendix B.
As expected, the co-product of the twisted Yangian generator $\tilde{\mathbb{R}}_{2}^{\enskip1}$ coincides
with the conserved  charge $\tilde{\mathbb{Q}}$ of \cite{Ahn1}.

Requiring that the reflection matrix respect the co-products (\ref{Y(g,h)}),
\begin{equation}
K(p)\,\Delta(\tilde{\mathbb{J}}^{p})-\Delta(\tilde{\mathbb{J}}^{p})\,K(p)=0,
\end{equation}
 one finds the reflection matrix $K^{B}$ coefficients to be
\begin{align}
 & k_{1}(p)=1,&& k_{2}(p)=-\frac{y^{-}+y^{-}\left(y^{+}\right)^{2}}{y^{+}+y^{-}\left(y^{+}\right)^{2}}, && k_{3}(p)=\frac{y^{-}}{y^{+}},\nonumber \\
 & k_{4}(p)=\frac{\left(1+\left(y^{-}\right)^{2}\right)y^{+}}{y^{-}\left(1+y^{-}y^{+}\right)}\frac{\tilde{\eta}^{2}}{\eta^{2}}, && k_{5}(p)=\frac{\tilde{\eta}}{\eta}, && k_{6}(p)=-\frac{\tilde{\eta}}{\eta},\nonumber \\
 & k_{7}(p)=\frac{i\tilde{\eta}^{2}}{\zeta\left(1+y^{-}y^{+}\right)}, && k_{8}(p)=\frac{i\zeta}{\left(1+y^{-}y^{+}\right)\eta^{2}}.
\end{align}
It is worthwhile to note that, from the scattering-theory point of
view, scattering from the $Y=0$ giant graviton is identical to that
from the $Y=0$ $D7$ brane \cite{Young1}. Consequently the reflection
matrices and boundary Yangian are the same.

\section{Discussion}

In this paper we have constructed a boundary remnant of Yangian
symmetry for the $Y=0$ giant graviton, as a special case of the generalized twisted Yangian
boundary symmetries $\mbox{Y}(\mathfrak{g},\mathfrak{h})$
of \cite{MacKay2}.

We have shown that the reflection matrices of the $Y=0$ giant graviton,
which preserves only an $\mathfrak{su}\left(2|1\right)$ subalgebra
of the centrally-extended $\mathfrak{psu}\left(2|2\right)$ algebra
and has no boundary degrees of freedom, respect a boundary remnant
$\mbox{Y}(\mathfrak{g},\mathfrak{h})$ of the bulk Yangian
symmetry $\mbox{Y}(\mathfrak{g})$ \cite{Beisert4} which extends
the results of \cite{Ahn1}: the conserved charge constructed in \cite{Ahn1}
is a generator of the twisted boundary Yangian $\mbox{Y}(\mathfrak{g},\mathfrak{h})$
we have constructed. Furthermore, from the scattering theory point
of view, the reflection from the $Y=0$ giant graviton is equivalent
to reflection from the $Y=0$ $D7$ brane \cite{Young1} and the corresponding
reflection matrices and Yangian symmetry are the same.

We have also considered the Yangian symmetry of a boundary which
preserves an $\mathfrak{su}\left(1|2\right)$ subalgebra of the centrally-extended
$\mathfrak{su}\left(2|2\right)$ algebra, with no boundary degrees of
freedom. We showed that this leads to a diagonal reflection
matrix for all $l$-magnon bound states and possesses a boundary Yangian.
However, this hidden symmetry is redundant, because
the reflection matrices are fully determined by the Lie symmetry algebra alone.
The meaning of this in the AdS/CFT correspondence is not clear; we present it
(in an appendix) merely as a nice example in
a contrast to the $Y=0$ giant graviton. One could try to consider
a boundary identical to the $Z=0$ giant graviton but with no boundary
degrees of freedom, but it is easy to check that this kind of configuration
is ruled out by the $\mathfrak{su}\left(2|2\right)$ algebra.

We have thus made some progress towards a better general understanding
of boundary symmetry and boundary states in AdS/CFT. There is one
further case of boundary symmetry which is not present in the case
of the giant graviton, but is present in the left factor of the reflection
from the $Z=0$ $D7$ brane \cite{Young1}. This preserves neither
supersymmetries nor boundary degrees of freedom, and the preserved
algebra is not part of a symmetric pair. A very similar reflection structure
was addressed in \cite{Nepomechie1}, and we hope to understand this
reflection problem in near future.

It is worth noting that the fundamental $S$-matrix possesses a secret
symmetry \cite{Torrielli4} which emerges from the exceptional superalgebra $\mathfrak{d}\left(2,1;\varepsilon\right)$
in the limit $\varepsilon\rightarrow0$ \cite{Matsumoto1}. We expect the
boundary Yangian to be rich in such secrets too.

\textbf{Acknowledgments}: We thank Alessandro Torrielli and Charles Young for 
comments and for reading the manuscript, and the UK EPSRC for funding
under grant EP/H000054/1.


\appendix

\section{Yangian symmetry of the $\mathfrak{su}(1|2)$ boundary}

We consider a spin chain ending on a boundary which preserves only
a $\mathfrak{h=su}\left(1|2\right)$ subalgebra of the bulk symmetry,
and which has no degrees of freedom attached to the end of the spin chain.
The commutation relations of $\mathfrak{su}\left(1|2\right)$ are inherited
from the (\ref{psu(2|2)_algebra}) by dropping the generators
with fermionic indices $\alpha,\,\beta,\,\gamma,\;...=4$; thus the
surviving generators are $\mathbb{R}_{a}^{\enskip b},$ $\mathbb{L}_{3}^{\enskip3}\equiv-\mathbb{L}_{4}^{\enskip4}$,
$\mathbb{Q}_{3}^{\enskip a}$, $\mathbb{G}_{a}^{\enskip3}$ and $\mathbb{H}$.
It is straightforward to check that the subalgebra $\mathfrak{h=su}\left(1|2\right)$
and subset $\mathfrak{m}=\mathfrak{psu}\left(2|2\right)\ltimes\mathbb{R}^{3}/\mathfrak{su}\left(1|2\right)$
consisting of generators $\mathbb{L}_{3}^{\enskip4},\;\mathbb{L}_{4}^{\enskip3},\;\mathbb{Q}_{4}^{\enskip a},\;\mathbb{G}_{a}^{\enskip4}$
and central charges $\mathbb{C},\;\mathbb{C}^{\dagger}$ form a symmetric
pair (\ref{symmetric_pair}) by considering the commutation relations in the
(\ref{psu(2|2)_algebra}). Thus the theory should possess a boundary
Yangian with the same structure as (\ref{Y(g,h)}). Using the general
prescription (\ref{twist}) one finds the non-trivial part of the
boundary Yangian co-products to be
\begin{align}
\Delta\tilde{\mathbb{L}}_{3}^{\enskip4} & =\left(\hat{\mathbb{L}}_{3}^{\enskip4}-\frac{1}{2}\mathbb{L}_{3}^{\enskip4}\,\mathbb{L}_{4}^{\enskip4}+\frac{1}{2}\mathbb{L}_{3}^{\enskip4}\,\mathbb{L}_{3}^{\enskip3}+\frac{1}{2}\mathbb{G}_{c}^{\enskip4}\,\mathbb{Q}_{3}^{\enskip c}\right)\otimes1,\nonumber \\
\Delta\tilde{\mathbb{L}}_{4}^{\enskip3} & =\left(\hat{\mathbb{L}}_{4}^{\enskip3}-\frac{1}{2}\mathbb{L}_{4}^{\enskip3}\mathbb{L}_{3}^{\enskip3}+\frac{1}{2}\mathbb{L}_{4}^{\enskip3}\,\mathbb{L}_{4}^{\enskip4}+\frac{1}{2}\mathbb{Q}_{4}^{\enskip c}\,\mathbb{G}_{c}^{\enskip3}\right)\otimes1,\nonumber \\
\Delta\tilde{\mathbb{Q}}_{4}^{\enskip a} & =\left(\hat{\mathbb{Q}}_{4}^{\enskip a}+\frac{1}{2}\mathbb{Q}_{4}^{\enskip c}\,\mathbb{R}_{c}^{\enskip a}+\frac{1}{2}\mathbb{Q}_{4}^{\enskip a}\,\mathbb{L}_{4}^{\enskip4}-\frac{1}{2}\mathbb{L}_{4}^{\enskip3}\,\mathbb{Q}_{3}^{\enskip a}+\frac{1}{4}\mathbb{Q}_{4}^{\enskip a}\,\mathbb{H}-\frac{1}{2}\varepsilon^{ad}\mathbb{C}\,\mathbb{G}_{d}^{\enskip3}\right)\otimes1,\nonumber \\
\Delta\tilde{\mathbb{G}}_{a}^{\enskip4} & =\left(\hat{\mathbb{G}}_{a}^{\enskip4}-\frac{1}{2}\mathbb{G}_{c}^{\enskip4}\,\mathbb{R}_{a}^{\enskip c}-\frac{1}{2}\mathbb{G}_{a}^{\enskip4}\,\mathbb{L}_{4}^{\enskip4}+\frac{1}{2}\mathbb{L}_{3}^{\enskip4}\,\mathbb{G}_{a}^{\enskip3}-\frac{1}{4}\mathbb{G}_{a}^{\enskip\alpha},\mathbb{H}+\frac{1}{2}\varepsilon_{ac}\mathbb{C}^{\dagger}\,\mathbb{Q}_{3}^{\enskip c}\right)\otimes1,\nonumber \\
\Delta\tilde{\mathbb{C}} & =\left(\hat{\mathbb{C}}+\frac{1}{2}\mathbb{C}\,\mathbb{H}\right)\otimes1,\nonumber \\
\Delta\mathbb{\tilde{C}}^{\dagger} & =\left(\hat{\mathbb{C}}^{\dagger}-\frac{1}{2}\mathbb{C}^{\dagger}\,\mathbb{H}\right)\otimes1.\label{Y(g,h)_2}
\end{align}
Once again, the full expressions of the co-products are presented in the appendix B.

We shall construct a scattering theory in the same way as was done in
 subsection 2.2 for the $Y=0$ giant graviton. First, we introduce
a boundary vacuum state $\left|0_{B}\right\rangle$ and a corresponding
trivial vector space $\mathcal{V}\left(0\right)$ which is annihilated
by all $\mathfrak{su}\left(1|2\right)$ generators. This construction
leads to the superspace $K$-matrix for the reflection of bulk magnons
from the boundary vacuum state as an operator acting on the tensor
product\begin{equation}
K(p):\quad\mathcal{V}^{M}(p,\zeta)\otimes\mathcal{V}_B(0)\rightarrow\mathcal{V}^{M}(-p,\zeta)\otimes\mathcal{V}_B(0),
\end{equation}
 where the reflection matrix is defined as a differential operator
\begin{equation}
K(p)=\sum_{i}k_{i}(p)\,\Lambda_{i}
\end{equation}
 acting on the superspace. In the case of the reflection of fundamental
states, following the similar considerations as for the $\mathfrak{su}(2|1)$ case, 
the symmetry algebra implies that the reflection matrix $K^{A}$
is a diagonal matrix
\begin{equation}
K^{A}(p)=k_{1}(p)\,\omega_{a}\frac{\partial}{\partial\omega_{a}}+k_{2}(p)\,\theta_{3}\frac{\partial}{\partial\theta_{3}}+k_{3}(p)\,\theta_{4}\frac{\partial}{\partial\theta_{4}}.
\end{equation}
Then the boundary symmetry algebra fixes the reflection coefficients
(up to an overall factor) to be
\begin{equation}
k_{1}(p)=1,\qquad k_{2}(p)=\frac{\tilde{\eta}}{\eta},\qquad k_{3}(p)=-\frac{x^{+}}{x^{-}}\frac{\tilde{\eta}}{\eta}.
\end{equation}
 In the case of the reflection of two-magnon bound states, the most
general structure of the reflection matrix $K^{B}$ one may write
is\begin{equation}
K^{B}(p)=\sum_{i=1}^{6}k_{i}(p)\,\Lambda_{i},
\end{equation}
 where $\Lambda_{i}$ with $i=1,...\,,4$ are diagonal and $\Lambda_{5}$,
$\Lambda_{6}$ are off-diagonal differential operators
\begin{align}
& \Lambda_{1}=\frac{1}{2}\omega_{b}\omega_{a}\frac{\partial^{2}}{\partial\omega_{b}\partial\omega_{a}},
&& \Lambda_{2}=\omega_{a}\theta_{3}\frac{\partial^{2}}{\partial\omega_{a}\partial\theta_{3}},
&& \Lambda_{3}=\omega_{a}\theta_{4}\frac{\partial^{2}}{\partial\omega_{a}\partial\theta_{4}},\nonumber \\
& \Lambda_{4}=\theta_{3}\theta_{4}\frac{\partial^{2}}{\partial\theta_{4}\partial\theta_{3}},
&& \Lambda_{5}=\theta_{3}\theta_{4}\frac{\partial^{2}}{\partial\omega_{2}\partial\omega_{1}},
&& \Lambda_{6}=\omega_{1}\omega_{2}\frac{\partial^{2}}{\partial\theta_{4}\partial\theta_{3}}.
\end{align}
However, the off-diagonal reflection channels are forbidden by the boundary symmetry.
It is easy to see this by considering the invariance of
the $K$-matrix under the $R$ symmetry generator $\mathbb{R}_{2}^{\;1}$
\begin{equation}
\mathbb{R}_{2}^{\;1}\, K^{B}\,\omega_{1}\omega_{1}=2\, k_{1}\,\omega_{1}\omega_{2}, \qquad
K^{B}\,\mathbb{R}_{2}^{\;1}\,\omega_{1}\omega_{1}=2\, k_{1}\,\omega_{1}\omega_{2}+2\, k_{5}\,\theta_{3}\theta_{4},
\end{equation}
leading to $k_{5}=0$, and
\begin{equation} \mathbb{R}_{2}^{\;1}\, K^{B}\,\theta_{3}\theta_{4}=k_{6}\,\omega_{2}\omega_{2}, \qquad
K^{B}\,\mathbb{R}_{2}^{\;1}\,\theta_{3}\theta_{4}=0,
\end{equation}
leading to $k_{6}=0$. This is a general feature for the reflection
of any $l$-magnon bound states. Hence the general $l$-magnon reflection
matrix \begin{equation}
K^{l}(p)=\sum_{i=1}^{4}k_{i}(p)\,\Lambda_{i}
\end{equation}
 is a diagonal matrix with
\begin{align}
& \Lambda_{1}=\frac{1}{l!}\,\omega_{a}^{l}\frac{\partial^{l}}{\partial^{l}\omega_{a}},
&& \Lambda_{2}=\frac{1}{\left(l-1\right)!}\,\omega_{a}^{l-1}\theta_{3}\frac{\partial^{l}}{\partial\omega_{a}^{l-1}\partial\theta_{3}},\nonumber \\
& \Lambda_{3}=\frac{1}{\left(l-1\right)!}\,\omega_{a}^{l-1}\theta_{4}\frac{\partial^{l}}{\partial\omega_{a}^{l-1}\partial\theta_{4}},
&& \Lambda_{4}=\frac{1}{\left(l-2\right)!}\,\omega_{a}^{l-2}\theta_{3}\theta_{4}\frac{\partial^{l}}{\partial\omega_{a}^{l-2}\partial\theta_{4}\partial\theta_{3}}.
\end{align}
The boundary symmetry algebra fixes the reflection coefficients uniquely
(up to an overall factor) without need of the boundary Yangian symmetry.
They are
\begin{equation}
k_{1}(p)=1,\qquad k_{2}(p)=\frac{\tilde{\eta}}{\eta},
\qquad k_{3}(p)=-\frac{y^{+}}{y^{-}}\frac{\tilde{\eta}}{\eta},
\qquad k_{4}(p)=-\frac{y^{+}}{y^{-}}\frac{\tilde{\eta}^{2}}{\eta^{2}},
\end{equation}
here $y^{\pm}$ are the spectral parameters of the $l$-magnon bound
state. On the other hand one could have used the boundary 
Yangian (\ref{Y(g,h)_2}) instead, leading to the same results.

This case is somewhat similar to the left factor of the reflection
from the $Z=0$ $D7$ brane \cite{Young1,Regelskis1}, which preserves neither
supersymmetries nor boundary degrees of freedom and allows only
diagonal reflection matrices $K^{l}$ identical for any $l\geq2$.
There, the absence of supersymmetries preserved by the boundary
required that the boundary Yang-Baxter equation be used to determine the
reflection matrices.

\section{The co-products of the boundary Yangian $Y(\mathfrak{g},\mathfrak{h})$}

The complete expressions of the co-products of the boundary twisted 
Yangian $\mathrm{Y}(\mathfrak{g},\mathfrak{h})$ defined by the
algebra $\mathfrak{g}=\mathfrak{psu}\left(2|2\right)\ltimes\mathbb{R}^{3}$
and subalgebra $\mathfrak{h=su}\left(2|1\right)$ preserved on the
boundary are
\begin{eqnarray}
\Delta\tilde{\mathbb{R}}_{1}^{\enskip2} & = & \left(\hat{\mathbb{R}}_{1}^{\enskip2}+\frac{1}{2}\mathbb{R}_{1}^{\enskip2}\,\mathbb{R}_{1}^{\enskip1}-\frac{1}{2}\mathbb{R}_{1}^{\enskip2}\,\mathbb{R}_{2}^{\enskip2}-\frac{1}{2}\mathbb{Q}_{\gamma}^{\enskip2}\,\mathbb{G}_{1}^{\enskip\gamma}\right)\otimes1\nonumber \\
 &  & +1\otimes\left(\hat{\mathbb{R}}_{1}^{\enskip2}+\frac{1}{2}\mathbb{R}_{1}^{\enskip2}\,\mathbb{R}_{1}^{\enskip1}-\frac{1}{2}\mathbb{R}_{1}^{\enskip2}\,\mathbb{R}_{2}^{\enskip2}-\frac{1}{2}\mathbb{Q}_{\gamma}^{\enskip2}\,\mathbb{G}_{1}^{\enskip\gamma}\right)\nonumber \\
 &  & +\mathbb{R}_{1}^{\enskip2}\otimes\mathbb{R}_{1}^{\enskip1}-\mathbb{R}_{1}^{\enskip2}\otimes\mathbb{R}_{2}^{\enskip2}-\mathbb{Q}_{\gamma}^{\enskip2}\otimes\mathbb{G}_{1}^{\enskip\gamma},\nonumber \\
\Delta\tilde{\mathbb{R}}_{2}^{\enskip1} & = & \left(\hat{\mathbb{R}}_{2}^{\enskip1}+\frac{1}{2}\mathbb{R}_{2}^{\enskip1}\,\mathbb{R}_{1}^{\enskip1}-\frac{1}{2}\mathbb{R}_{2}^{\enskip1}\,\mathbb{R}_{2}^{\enskip2}-\frac{1}{2}\mathbb{G}_{2}^{\enskip\gamma}\,\mathbb{Q}_{\gamma}^{\enskip1}\right)\otimes1\nonumber \\
 &  & +1\otimes\left(\hat{\mathbb{R}}_{2}^{\enskip1}+\frac{1}{2}\mathbb{R}_{2}^{\enskip1}\,\mathbb{R}_{1}^{\enskip1}-\frac{1}{2}\mathbb{R}_{2}^{\enskip1}\,\mathbb{R}_{2}^{\enskip2}-\frac{1}{2}\mathbb{G}_{2}^{\enskip\gamma}\,\mathbb{Q}_{\gamma}^{\enskip1}\right)\nonumber \\
 &  & +\mathbb{R}_{2}^{\enskip1}\otimes\mathbb{R}_{1}^{\enskip1}-\mathbb{R}_{2}^{\enskip1}\otimes\mathbb{R}_{2}^{\enskip2}-\mathbb{G}_{2}^{\enskip\gamma}\otimes\mathbb{Q}_{\gamma}^{\enskip1},\nonumber \\
\Delta\tilde{\mathbb{Q}}_{\alpha}^{\enskip2} & = & \left(\hat{\mathbb{Q}}_{\alpha}^{\enskip2}+\frac{1}{2}\mathbb{Q}_{\alpha}^{\enskip2}\,\mathbb{R}_{2}^{\enskip2}-\frac{1}{2}\mathbb{R}_{1}^{\enskip2}\,\mathbb{Q}_{\alpha}^{\enskip1}+\frac{1}{2}\mathbb{Q}_{\gamma}^{\enskip2}\,\mathbb{L}_{\alpha}^{\enskip\gamma}+\frac{1}{4}\mathbb{Q}_{\alpha}^{\enskip2}\,\mathbb{H}-\frac{1}{2}\varepsilon_{\alpha\gamma}\mathbb{C}\,\mathbb{G}_{1}^{\enskip\gamma}\right)\otimes1\nonumber \\
 &  & +1\otimes\left(\hat{\mathbb{Q}}_{\alpha}^{\enskip2}+\frac{1}{2}\mathbb{Q}_{\alpha}^{\enskip2}\,\mathbb{R}_{2}^{\enskip2}-\frac{1}{2}\mathbb{R}_{1}^{\enskip2}\,\mathbb{Q}_{\alpha}^{\enskip1}+\frac{1}{2}\mathbb{Q}_{\gamma}^{\enskip2}\,\mathbb{L}_{\alpha}^{\enskip\gamma}+\frac{1}{4}\mathbb{Q}_{\alpha}^{\enskip2}\,\mathbb{H}-\frac{1}{2}\varepsilon_{\alpha\gamma}\mathbb{C}\,\mathbb{G}_{1}^{\enskip\gamma}\right)\nonumber \\
 &  & +\mathbb{Q}_{\alpha}^{\enskip2}\otimes\mathbb{R}_{2}^{\enskip2}-\mathbb{R}_{1}^{\enskip2}\otimes\mathbb{Q}_{\alpha}^{\enskip1}+\mathbb{Q}_{\gamma}^{\enskip2}\otimes\mathbb{L}_{\alpha}^{\enskip\gamma}+\frac{1}{2}\mathbb{Q}_{\alpha}^{\enskip2}\otimes\mathbb{H}-\varepsilon_{\alpha\gamma}\mathbb{C}\otimes\mathbb{G}_{1}^{\enskip\gamma},\nonumber
\end{eqnarray}
\begin{eqnarray}
\Delta\tilde{\mathbb{G}}_{2}^{\enskip\alpha} & = & \left(\hat{\mathbb{G}}_{2}^{\enskip\alpha}-\frac{1}{2}\mathbb{G}_{2}^{\enskip\alpha}\,\mathbb{R}_{2}^{\enskip2}+\frac{1}{2}\mathbb{R}_{2}^{\enskip1}\,\mathbb{G}_{1}^{\enskip\alpha}-\frac{1}{2}\mathbb{G}_{2}^{\enskip\gamma}\,\mathbb{L}_{\gamma}^{\enskip\alpha}-\frac{1}{4}\mathbb{G}_{2}^{\enskip\alpha}\,\mathbb{H}+\frac{1}{2}\varepsilon^{\alpha\gamma}\mathbb{C}^{\dagger}\,\mathbb{Q}_{\gamma}^{\enskip1}\right)\otimes1\nonumber \\
 &  & +1\otimes\left(\hat{\mathbb{G}}_{2}^{\enskip\alpha}-\frac{1}{2}\mathbb{G}_{2}^{\enskip\alpha}\,\mathbb{R}_{2}^{\enskip2}+\frac{1}{2}\mathbb{R}_{2}^{\enskip1}\,\mathbb{G}_{1}^{\enskip\alpha}-\frac{1}{2}\mathbb{G}_{2}^{\enskip\gamma}\,\mathbb{L}_{\gamma}^{\enskip\alpha}-\frac{1}{4}\mathbb{G}_{2}^{\enskip\alpha}\,\mathbb{H}+\frac{1}{2}\varepsilon^{\alpha\gamma}\mathbb{C}^{\dagger}\,\mathbb{Q}_{\gamma}^{\enskip1}\right)\nonumber \\
 &  & -\mathbb{G}_{2}^{\enskip\alpha}\otimes\mathbb{R}_{2}^{\enskip2}+\mathbb{R}_{2}^{\enskip1}\otimes\mathbb{G}_{1}^{\enskip\alpha}-\mathbb{G}_{2}^{\enskip\gamma}\otimes\mathbb{L}_{\gamma}^{\enskip\alpha}-\frac{1}{2}\mathbb{G}_{2}^{\enskip\alpha}\otimes\mathbb{H}+\varepsilon^{\alpha\gamma}\mathbb{C}^{\dagger}\otimes\mathbb{Q}_{\gamma}^{\enskip1},\nonumber \\
\Delta\tilde{\mathbb{C}} & = & \left(\hat{\mathbb{C}}+\frac{1}{2}\mathbb{C}\,\mathbb{H}\right)\otimes1+1\otimes\left(\hat{\mathbb{C}}+\frac{1}{2}\mathbb{C}\,\mathbb{H}\right)+\mathbb{C}\otimes\mathbb{H},\nonumber \\
\Delta\mathbb{\tilde{C}}^{\dagger} & = & \left(\hat{\mathbb{C}}^{\dagger}-\frac{1}{2}\mathbb{C}^{\dagger}\,\mathbb{H}\right)
\otimes1+1\otimes\left(\hat{\mathbb{C}}^{\dagger}-\frac{1}{2}\mathbb{C}^{\dagger}\,
\mathbb{H}\right)-\mathbb{C}^{\dagger}\otimes\mathbb{H},
\end{eqnarray}
where one can observe the co-ideal property explicitly.

For the second, toy case, where $\mathfrak{g}=\mathfrak{psu}\left(2|2\right)\ltimes\mathbb{R}^{3}$
and $\mathfrak{h=su}\left(1|2\right)$ is the subalgebra preserved on the
boundary, the complete expressions of the co-products are
\begin{eqnarray}
\Delta\tilde{\mathbb{L}}_{3}^{\enskip4} & = & \left(\hat{\mathbb{L}}_{3}^{\enskip4}-\frac{1}{2}\mathbb{L}_{3}^{\enskip4}\,\mathbb{L}_{4}^{\enskip4}+\frac{1}{2}\mathbb{L}_{3}^{\enskip4}\,\mathbb{L}_{3}^{\enskip3}+\frac{1}{2}\mathbb{G}_{c}^{\enskip4}\,\mathbb{Q}_{3}^{\enskip c}\right)\otimes1\nonumber \\
 &  & +1\otimes\left(\hat{\mathbb{L}}_{3}^{\enskip4}-\frac{1}{2}\mathbb{L}_{3}^{\enskip4}\,\mathbb{L}_{4}^{\enskip4}+\frac{1}{2}\mathbb{L}_{3}^{\enskip4}\,\mathbb{L}_{3}^{\enskip3}+\frac{1}{2}\mathbb{G}_{c}^{\enskip4}\,\mathbb{Q}_{3}^{\enskip c}\right)\nonumber \\
 &  & -\mathbb{L}_{3}^{\enskip4}\otimes\mathbb{L}_{4}^{\enskip4}+\mathbb{L}_{3}^{\enskip4}\otimes\mathbb{L}_{3}^{\enskip3}+\mathbb{G}_{c}^{\enskip4}\otimes\mathbb{Q}_{3}^{\enskip c},\nonumber \\
\Delta\tilde{\mathbb{L}}_{4}^{\enskip3} & = & \left(\hat{\mathbb{L}}_{4}^{\enskip3}-\frac{1}{2}\mathbb{L}_{4}^{\enskip3}\,\mathbb{L}_{3}^{\enskip3}+\frac{1}{2}\mathbb{L}_{4}^{\enskip3}\,\mathbb{L}_{4}^{\enskip4}+\frac{1}{2}\mathbb{Q}_{4}^{\enskip c}\,\mathbb{G}_{c}^{\enskip3}\right)\otimes1\nonumber \\
 &  & +1\otimes\left(\hat{\mathbb{L}}_{4}^{\enskip3}-\frac{1}{2}\mathbb{L}_{4}^{\enskip3}\,\mathbb{L}_{3}^{\enskip3}+\frac{1}{2}\mathbb{L}_{4}^{\enskip3}\,\mathbb{L}_{4}^{\enskip4}+\frac{1}{2}\mathbb{Q}_{4}^{\enskip c}\,\mathbb{G}_{c}^{\enskip3}\right)\nonumber \\
 &  & -\mathbb{L}_{4}^{\enskip3}\otimes\mathbb{L}_{3}^{\enskip3}+\mathbb{L}_{4}^{\enskip3}\otimes\mathbb{L}_{4}^{\enskip4}+\mathbb{Q}_{4}^{\enskip c}\otimes\mathbb{G}_{c}^{\enskip3},\nonumber \\
\Delta\tilde{\mathbb{Q}}_{4}^{\enskip a} & = & \left(\hat{\mathbb{Q}}_{4}^{\enskip a}+\frac{1}{2}\mathbb{Q}_{4}^{\enskip c}\,\mathbb{R}_{c}^{\enskip a}+\frac{1}{2}\mathbb{Q}_{4}^{\enskip a}\,\mathbb{L}_{4}^{\enskip4}-\frac{1}{2}\mathbb{L}_{4}^{\enskip3}\,\mathbb{Q}_{3}^{\enskip a}+\frac{1}{4}\mathbb{Q}_{4}^{\enskip a}\,\mathbb{H}-\frac{1}{2}\varepsilon^{ad}\mathbb{C}\,\mathbb{G}_{d}^{\enskip3}\right)\otimes1\nonumber \\
 &  & +1\otimes\left(\hat{\mathbb{Q}}_{4}^{\enskip a}+\frac{1}{2}\mathbb{Q}_{4}^{\enskip c}\,\mathbb{R}_{c}^{\enskip a}+\frac{1}{2}\mathbb{Q}_{4}^{\enskip a}\,\mathbb{L}_{4}^{\enskip4}-\frac{1}{2}\mathbb{L}_{4}^{\enskip3}\,\mathbb{Q}_{3}^{\enskip a}+\frac{1}{4}\mathbb{Q}_{4}^{\enskip a}\,\mathbb{H}-\frac{1}{2}\varepsilon^{ad}\mathbb{C}\,\mathbb{G}_{d}^{\enskip3}\right)\nonumber \\
 &  & +\mathbb{Q}_{4}^{\enskip c}\otimes\mathbb{R}_{c}^{\enskip a}+\mathbb{Q}_{4}^{\enskip a}\otimes\mathbb{L}_{4}^{\enskip4}-\mathbb{L}_{4}^{\enskip3}\otimes\mathbb{Q}_{3}^{\enskip a}+\frac{1}{2}\mathbb{Q}_{4}^{\enskip a}\otimes\mathbb{H}-\varepsilon^{ad}\mathbb{C}\otimes\mathbb{G}_{d}^{\enskip3},\nonumber \\
\Delta\tilde{\mathbb{G}}_{a}^{\enskip4} & = & \left(\hat{\mathbb{G}}_{a}^{\enskip4}-\frac{1}{2}\mathbb{G}_{c}^{\enskip4}\,\mathbb{R}_{a}^{\enskip c}-\frac{1}{2}\mathbb{G}_{a}^{\enskip4}\,\mathbb{L}_{4}^{\enskip4}+\frac{1}{2}\mathbb{L}_{3}^{\enskip4}\,\mathbb{G}_{a}^{\enskip3}-\frac{1}{4}\mathbb{G}_{a}^{\enskip\alpha},\mathbb{H}+\frac{1}{2}\varepsilon_{ac}\mathbb{C}^{\dagger}\,\mathbb{Q}_{3}^{\enskip c}\right)\otimes1\nonumber \\
 &  & +1\otimes\left(\hat{\mathbb{G}}_{a}^{\enskip4}-\frac{1}{2}\mathbb{G}_{c}^{\enskip4}\,\mathbb{R}_{a}^{\enskip c}-\frac{1}{2}\mathbb{G}_{a}^{\enskip4}\,\mathbb{L}_{4}^{\enskip4}+\frac{1}{2}\mathbb{L}_{3}^{\enskip4}\,\mathbb{G}_{a}^{\enskip3}-\frac{1}{4}\mathbb{G}_{a}^{\enskip\alpha}\,\mathbb{H}+\frac{1}{2}\varepsilon_{ac}\mathbb{C}^{\dagger}\,\mathbb{Q}_{3}^{\enskip c}\right)\nonumber \\
 &  & -\mathbb{G}_{c}^{\enskip4}\otimes\mathbb{R}_{a}^{\enskip c}-\mathbb{G}_{a}^{\enskip4}\otimes\mathbb{L}_{4}^{\enskip4}+\mathbb{L}_{3}^{\enskip4}\otimes\mathbb{G}_{a}^{\enskip3}-\frac{1}{2}\mathbb{G}_{a}^{\enskip\alpha}\otimes\mathbb{H}+\varepsilon_{ac}\mathbb{C}^{\dagger}\otimes\mathbb{Q}_{3}^{\enskip c},\nonumber \\
\Delta\tilde{\mathbb{C}} & = & \left(\hat{\mathbb{C}}+\frac{1}{2}\mathbb{C}\,\mathbb{H}\right)\otimes1+1\otimes\left(\hat{\mathbb{C}}+\frac{1}{2}\mathbb{C}\,\mathbb{H}\right)+\mathbb{C}\otimes\mathbb{H},\nonumber \\
\Delta\mathbb{\tilde{C}}^{\dagger} & = & \left(\hat{\mathbb{C}}^{\dagger}-\frac{1}{2}\mathbb{C}^{\dagger}\,\mathbb{H}\right)\otimes1+1\otimes\left(\hat{\mathbb{C}}^{\dagger}-\frac{1}{2}\mathbb{C}^{\dagger}\,\mathbb{H}\right)-\mathbb{C}^{\dagger}\otimes\mathbb{H}.
\end{eqnarray}

All generators in the right factors of the expressions above annihilate
the boundary vacuum state $\left|0_B\right\rangle$; hence the
reflection matrices do not implicitly depend on them.
However, we do not expect these full co-products to be valid for boundaries
\textit{with}  degrees of freedom, for the presence of the latter would change
the boundary algebra structure. Further considerations are beyond
the scope of the present work, but are the subject of current investigations.\\


\begin{thebibliography}{30}
\bibitem{Zarembo1}J. A. Minahan, K. Zarembo, \textit{The Bethe-Ansatz
for N=4 Super Yang-Mills}, JHEP 0303 (2003) 013, {[}\hepth{0212208}{]}.

\bibitem{Bena}I. Bena, J. Polchinski and R. Roiban, \textit{Hidden
symmetries of the $AdS_{5}\times S^{5}$ super-string}, Phys. Rev.
D 69 (2004) 046002 {[}\hepth{0305116}{]}.

\bibitem{Dolan}L. Dolan, C.R. Nappi, E. Witten, \textit{Yangian Symmetry
in D=4 Superconformal Yang-Mills Theory}, {[}\hepth{0401243}{]}.

\bibitem{Maldacena1}J. M. Maldacena, \textit{The Large N Limit of
Superconformal Field Theories and Supergravity}, Adv.Theor.Math.Phys.2:231-252,
(1998) {[}\hepth{9711200}{]}.

\bibitem{Beisert2}N. Beisert, \textit{The $su(2|2)$ Dynamic S-Matrix},
Adv.Theor.Math.Phys.12:945, (2008) {[}\hepth{0511082v4}{]}.

\bibitem{Arutyunov2}G. Arutyunov and S. Frolov, \textit{Foundations
of the $AdS_{5}\times S^{5}$ Superstring. Part I}, {[}\arXivid{0901.4937}{]}.

\bibitem{Staudacher1}M. Staudacher, \textit{The Factorized S-Matrix
of AdS/CFT}, JHEP 0505:054, (2005) {[}\hepth{0412188v1}{]}.

\bibitem{Beisert1}N. Beisert,\textit{ The Analytic Bethe Ansatz for
a Chain with Centrally Extended $su(2|2)$ Symmetry}, J.Stat.Mech.0701:P017,
(2007) {[}\nlin{SI}{0610017}{]}.


\bibitem{Dorey2}N. Dorey, \textit{Magnon Bound States and the AdS/CFT
Correspondence}, J.Phys.A39:13119-13128, (2006) {[}\hepth{0604175v2}{]}.

\bibitem{Dorey1}H. Yu Chen, N. Dorey, Keisuke Okamura, \textit{The
Asymptotic Spectrum of the $\mathcal{N}=4$ Super Yang-Mills Spin
Chain}, JHEP 0703 (2007) 005, {[}\hepth{0610295v1}{]}.

\bibitem{Arutyunov1}G. Arutyunov, S. Frolov, \textit{The $S$-matrix
of String Bound States}, Nucl.Phys.B804:90-143, (2008) {[}\arXivid{0803.4323}{]}.

\bibitem{Beisert4}N. Beisert, \textit{The S-Matrix of AdS/CFT and
Yangian Symmetry}, PoSSolvay:002, (2006) {[}\arXivid{0704.0400}{]}.

\bibitem{Zwiebel1}B. I. Zwiebel, \textit{Yangian Symmetry at Two
Loops for the $su(2|1)$ Sector of N=4 SYM}, J. Phys. A40 (2007) 1141-1152,
{[}\hepth{0610283}{]}.

\bibitem{Drinfeld1}V. Drinfeld, \textit{Hopf algebras and the quantum
Yang-Baxter equation}, Sov. Math. Dokl. 32 (1985) 254.

\bibitem{Bernard1}D. Bernard, \textit{An Introduction to Yangian
Symmetries}, Int. J. Mod. Phys. B7 (1993) 3517\textendash{}3530, {[}\hepth{9211133}{]}.

\bibitem{Ragoucy1}D. Arnaudon, A. Molev, E. Ragoucy, \textit{On the
R-matrix realization of Yangians and their representations}, Annales
Henri Poincare, 7 (2006), 1269-1325, {[}\Math{QA}{0511481v1}{]}.

\bibitem{deLeeuw1}M. de Leeuw, \textit{Bound States, Yangian Symmetry
and Classical r-matrix for the $AdS_{5}\times S^{5}$ Superstring},
JHEP 0806 (2008) 085, {[}\arXivid{0804.1047}{]}.

\bibitem{Arutyunov6}G. Arutyunov, M. de Leeuw, A. Torrielli , \textit{The
Bound State S-Matrix for $AdS_{5}\times S^{5}$ Superstring}, Nucl.
Phys. B819 (2009) 319-350, {[}\arXivid{0902.0183}{]}.

\bibitem{Ahn1}C. Ahn, R. I. Nepomechie, \textit{Yangian symmetry
and bound states in AdS/CFT boundary scattering}, JHEP 1005 (2010)
016, {[}\arXivid{1003.3361}{]}.

\bibitem{MacKay1}N. J. MacKay, \textit{Introduction to Yangian symmetry
in integrable field theory}, Int. J. Mod. Phys. A20 (2005) 7189\textendash{}7218,
{[}\hepth{0409183}{]}.

\bibitem{MacKay2}G. W. Delius, N. J. MacKay and B. J. Short, \textit{Boundary
remnant of Yangian symmetry and the structure of rational reflection
matrices}, Phys.Lett.B522:335-344, (2001) {[}\hepth{0109115}{]}.

\bibitem{Doikou1}A. Doikou, On boundary super algebras, J. Math.
Phys. 51:043509, 2010, {[}\arXivid{0910.1203}{]}.

\bibitem{Molev}A. I. Molev, Yangians and their applications, Handbook
of Algebra, Vol. 3, Elsevier, 907-959 (2003), {[}\Math{QA}{0211288}{]}.

\bibitem{Maldacena2}D. M. Hofman, J. Maldacena, \textit{Reflecting
magnons}, JHEP 0711 (2007) 050, {[}\arXivid{0708.2272}{]}.

\bibitem{Matsumoto1}T. Matsumoto, S. Moriyama, \textit{An Exceptional
Algebraic Origin of the AdS/CFT Yangian Symmetry}, JHEP 0804 (2008)
022, {[}\arXivid{0803.1212}{]}.

\bibitem{Spill1}F. Spill, A. Torrielli, \textit{On Drinfeld's second 
realization of the AdS/CFT su(2$\vert$2) Yangian}, J.Geom.Phys.59:489-502 
(2009), {[}\arXivid{0803.3194}{]}.

\bibitem{MacKay3}N. J. MacKay, B. J. Short, \textit{Boundary scattering,
symmetric spaces and the principal chiral model on the half line},
Comm. Math. Phys. 233:313-354, 2003, Erratum-ibid.245:425-428, 2004, {[}\hepth{0104212}{]}.

\bibitem{Young1}D. H. Correa, C. A. S. Young, \textit{Reflecting
magnons from D7 and D5 branes}, J.Phys.A41:455401, (2008) {[}\arXivid{0808.0452}{]}.

\bibitem{Nepomechie1}R. I. Nepomechie, E. Ragoucy, \textit{Analytical
Bethe ansatz for the open AdS/CFT SU(1$\vert$1) spin chain}, JHEP 0812
(2008) 025, {[}\arXivid{0810.5015}{]}.

\bibitem{Torrielli4}T. Matsumoto, S. Moriyama, A. Torrielli, \textit{A
Secret Symmetry of the AdS/CFT S-matrix}, JHEP 0709 (2007) 099, {[}\arXivid{0708.1285}{]}.

\bibitem{Regelskis1}N. MacKay, V. Regelskis, \textit{On the reflection
of magnon bound states}, JHEP 1008 (2010) 055, {[}\arXivid{1006.4102}{]}.

\end{thebibliography}
\end{document}